\documentclass[11pt]{article}
\usepackage{czerny,epsfig}

\def\Journal#1#2#3#4{{#1} {\bf #2}, #3 (#4)}

\def\apj{\em ApJ}
\def\aa{\em A\&A}
\def\mn{\em MNRAS}
\def\nat{\em Nature}
\def\phr{\em Physics Reports}
\def\apjs{\em ApJS}
\def\aj{\em AJ}
\def\araa{\em Ann. Rev. Astron. Astrophys.}

\def\be{\begin{equation}}
\def\ee{\end{equation}}
\def\bea{\begin{eqnarray}}
\def\eea{\end{eqnarray}}

\begin{document}
\vspace*{4cm}
\title{RADIO QUIET AGN}

\author{ B. CZERNY$^1$, R. GOOSMANN$^2$, V. KARAS$^3$, G. PONTI$^4$}

\address{$^1$Copernicus Astronomical Center, Bartycka 18, 00-716 Warsaw, 
   Poland\\
    $^2$Observatoire de Paris-Maudon, LUTH, Meudon, France\\
   $^3$Astronomical Institute, Academy of Sciences, Bocni II, 141 31 Prague, Czech Republic\\
   $^4$ ASF-CNR, Sezione di Bologna, via Gobetti 101, 40129 Bologna, Italy\\}

\maketitle\abstracts{
Active Galactic Nuclei are powered by accretion onto massive black holes. Although radio-quiet
objects are not as spectacular sources of very high energy photons as radio-loud ones this
class of objects also represents a challenge for modeling high energy processes close to
a black hole. Both a hot optically thin plasma and a cooler optically thick accretion disk are
usually thought to be present in the vicinity of a black hole although the details of the 
accretion flow are still
under discussion. The role of the disk seems to decrease with a drop in the Eddington ratio:
in sources like quasars and Narrow Line Seyfert 1 galaxies disk flow dominates while 
in Seyfert galaxies the disk retreats, and in sources like LINERS or Sgr A* 
a disk is most likely absent.
Shocks and reconnections are possibly taking place in an
inner hot flow and in the magnetic corona above the cold disk. Uncollimated outflow is also 
present and it may carry significant fraction of available mass and energy.  
}

\section{Introduction}
Active Galactic Nuclei (hereafter AGN) are the most powerful persistent and compact 
sources of radiation 
in the Universe. As such, they are clearly of interest from the point of view of high
energy processes. AGN activity is caused by accretion of material onto the massive black hole residing at 
the center of a host galaxy. Therefore, in those objects we can observe the behavior of the matter in a strong gravity field.  

Although the basic engine is always the same ---  extraction of gravitational energy of the infalling material ---  AGN are quite an inhomogeneous class of objects. The most 
important differentiating property is perhaps the radio loudness. A small fraction of 
those sources are strong radio emitters and display spectacular jets and/or radio extended radio lobes.  The majority, however, forms a population of radio 
quiet sources. The transition between the two populations is rather smooth, but 
nevertheless it is convenient to introduce two separate classes of objects, with 
a focus on strong relativistic jets as the main property of radio-loud sources. 
The customary border is set at the 5 GHz radio to B band flux ratio 
$ \log F_{5GHz}/\log F_B = 10$. The properties of the radio-loud objects are discussed by 
R. Moderski and A. Celotti (these proceedings). Here I will concentrate on the 
radio-quiet population and their relevance to high energy astrophysics.

\section{Basic facts}

AGN span a very broad range of luminosities. The brightest AGN (found among quasars)
have bolometric luminosities almost up to $\sim 10^{48}$ erg s$^{-1}$ \cite{woo}. They 
are mostly found among the high redshift objects. Nowadays, numerous quasars
are found up to $z \sim 6$ \cite{highz} as a result of massive surveys (e.g.
SDSS \cite{sdss}).
Nearby Seyfert galaxies are a few orders of magnitude fainter. The lower limit for a 
nuclear activity is 
unspecified since the determination of the weak nuclear activity is 
observationally difficult. However, it is now widely believed that all regular 
galaxies with bulges contain a supermassive black hole and must show some level of 
activity. In this sense we can also count SGR A* as an example of weak activity 
(see S. Nayakshin, these proceedings),
with its occasional flares  reaching up to the level of $10^{35}$ erg s$^{-1}$.

Black hole mass estimates indicate much narrower range, 
$\sim 10^6 -  10^{10} M_{\odot}$ \cite{wandel}, 
than the luminosity range
 which means that the Eddington ratio differs considerably between the sources. 

\subsection{Classification}

Radio-quiet AGN are divided into several classes, with two parameters
most plausibly underlying this classification scheme: inclination angle
and the Eddington ratio. 

Generally, all AGN are 
divided into type 1 and type 2 objects. Type 1 objects have very broad emission lines while 
type 2 objects show only narrow lines in their optical/UV spectra. It is generally accepted 
that class 2 is simply an obscured 
version of type 1 objects so we have no direct view of the nucleus in those sources.
This obscuration is due to the material predominantly located in the equatorial plane,
in a form of a dusty/molecular torus.
It can be noted, however, that there may be intrinsic type 2 sources among low 
Eddington ratio AGN \cite{tran}. It only means that classification based on the width 
of the hydrogen lines may not always represent well the characteristics of the 
central engine. In further text we will discuss only type 1 AGN.

Another classification, mostly historical, divides type 1 radio quiet AGN into quasars, 
Seyfert galaxies, Narrow Line Seyfert 1 galaxies and LINERS. 
The transition between these classes (connected mostly with 
the luminosity and the Eddington ratio) is smooth. Moreover, some quasars are also 
classified as NLS1 galaxies, 
and some LINERS perhaps are mostly starburst so the issue is confusing but we will 
preserve these subgroups to indicate the luminosity class of the discussed sources. Also
for bright quasars the transition to the corresponding 'Narrow Line Type 1' class happens at
much higher width ($\sim 4000$ km s$^{-1}$ \cite{bachev}) than for Seyfert 
galaxies ($\sim 2000$ km s$^{-1}$). This can be easily understood if the transition
happens at a fixed Eddington ratio and the difference
in black hole mass is taken into account (see formulae in \cite{BLR}). 

\subsection{Broad band spectra}

\begin{figure}
\psfig{figure=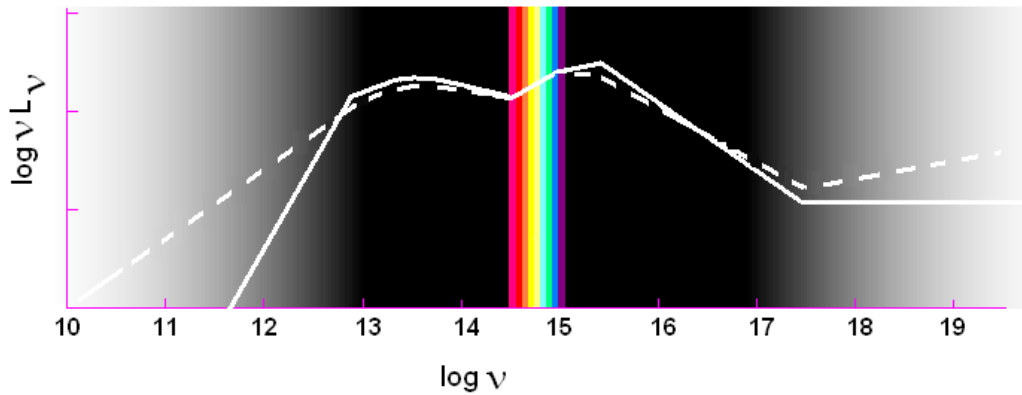,height=2.4 in}
\caption{Schematic view of the radiation spectrum of a bright radio quiet (continuous line) and radio-loud (dashed line) AGN.
\label{fig:spectrum}}
\end{figure}

The radiation spectrum of radio quiet AGN is very broad and span from radio to 
gamma band. The spectrum is best studied for high luminosity sources. A schematic 
view, representative for sources with high Eddington ratio (quasars and Narrow 
Line Seyfert 1 galaxies; $L/L_{Edd} \sim 1$) is shown in Fig.~\ref{fig:spectrum} \cite{elvis94}.
The spectrum usually peaks in the far UV/soft X-ray range, not accessible to the 
observations due to the Galactic extinction. However, the observational gap can be 
partially filled  by combining low redshift and high redshift sources into a single 
composite spectrum \cite{laor}. The extension of the spectra beyond 
100 keV is not well studied and we return to this point in Sect.~\ref{sect:gamma_spectra}.  

The spectra of AGN with lower Eddington ratio like Seyfert 1 galaxies 
($L/L_{Edd} \sim 0.01$) show less 
pronounced far UV peak and relatively stronger X-ray emission, and the X-ray spectrum 
is harder. The black holes accreting at the lowest rate ($L/L_{Edd} \sim 10^{-4}$) 
like those in LINERS or LLAGN do not show any strong UV 
emission while they are still relatively bright in X-rays. The broad band spectra of 
such objects are difficult to determine since the optical spectrum is in this case 
dominated by the host galaxy. Their X-ray spectra are as hard as in Seyfert 1 galaxies. 

Practically all radio-quiet AGN are rather variable \cite{markowitz,optical}. The shortest 
variability timescales (hundreds of seconds) are seen in X-ray band, in Seyfert galaxies. 
Optical/UV emission varies in timescales of days in those sources while in quasars 
monitoring over years is needed to see significant changes in optical band \cite{giveon}.  

\section{Radio-quiet AGN as sources of high energy photons and particles}

Radio-quiet AGN are not as obviously attractive from the point of view of the high 
energy physics as are radio-loud AGN. However, they also represent certain challenge 
for such studies. In order to show that we will discuss now the 
gamma-ray emission from these sources and the most plausible environment of its production.

\subsection{Extension of the AGN spectra into high energies}
\label{sect:gamma_spectra}

The direct signature of the importance of high energy processes in a given object is the presence of the gamma-ray emission.
Unfortunately,  spectral measurements going beyond 100 keV were performed only for a few sources, 
and the results were uncertain. Fits to the OSSE composite 
spectrum \cite{Zdzia-OSSE} gave the high
energy cut-off $120^{+220}_{-60}$ keV for Seyfert 1 and $130^{+220}_{-50}$ keV for Seyfert 2
galaxies (at such high energies the obscuration by the torus is relatively unimportant).
{\it Beppo-SAX} data, however, did not give such a strong constraints 
(cut-off energy $231^{+ \infty}_{-168}$ keV for Seyfert 1 galaxies, no cut-off 
for Seyfert 2 galaxies \cite{deluit}). Comparison of the model to the observed X-ray background
suggest that if the typical photon index is $\sim 1.9$ the spectrum extends up to $\sim 300$ 
keV \cite{background} while observations of high redshift quasars give results from 100 keV to 500 keV
and more, depending on the object and the adopted geometry \cite{sobo1,sobo2}.

Overall similarity of AGN and galactic black holes may suggest that sources with
rather hard spectra (photon index 1.9 or less) are thermal (electron temperature of order of 100 keV). Such a spectrum is well explained as a result 
of the Comptonization of the soft photons by a predominantly thermal plasma with the electron 
temperature around $10^9$ K. Ions may have much higher temperature, or may not be fully 
thermalized, but we have no direct methods of estimating their properties. Efficient 
thermalization is supposed to be achieved through synchrotron self-absorption \cite{ghisellini}.
Sources with steep spectra (photon index 2.0 and more) are likely to be significantly non-thermal, with a population of electrons having a power law distribution of energies, as expected in case of effective acceleration and ineffective thermalization.

Direct observation constraints should come from Astro-E2 and GLAST.

\subsection{accretion flow geometry}

\begin{figure}
\parbox{\textwidth}{
\parbox{0.5\textwidth}{
\epsfxsize=0.45\textwidth
\epsfbox[18 200 600 720]{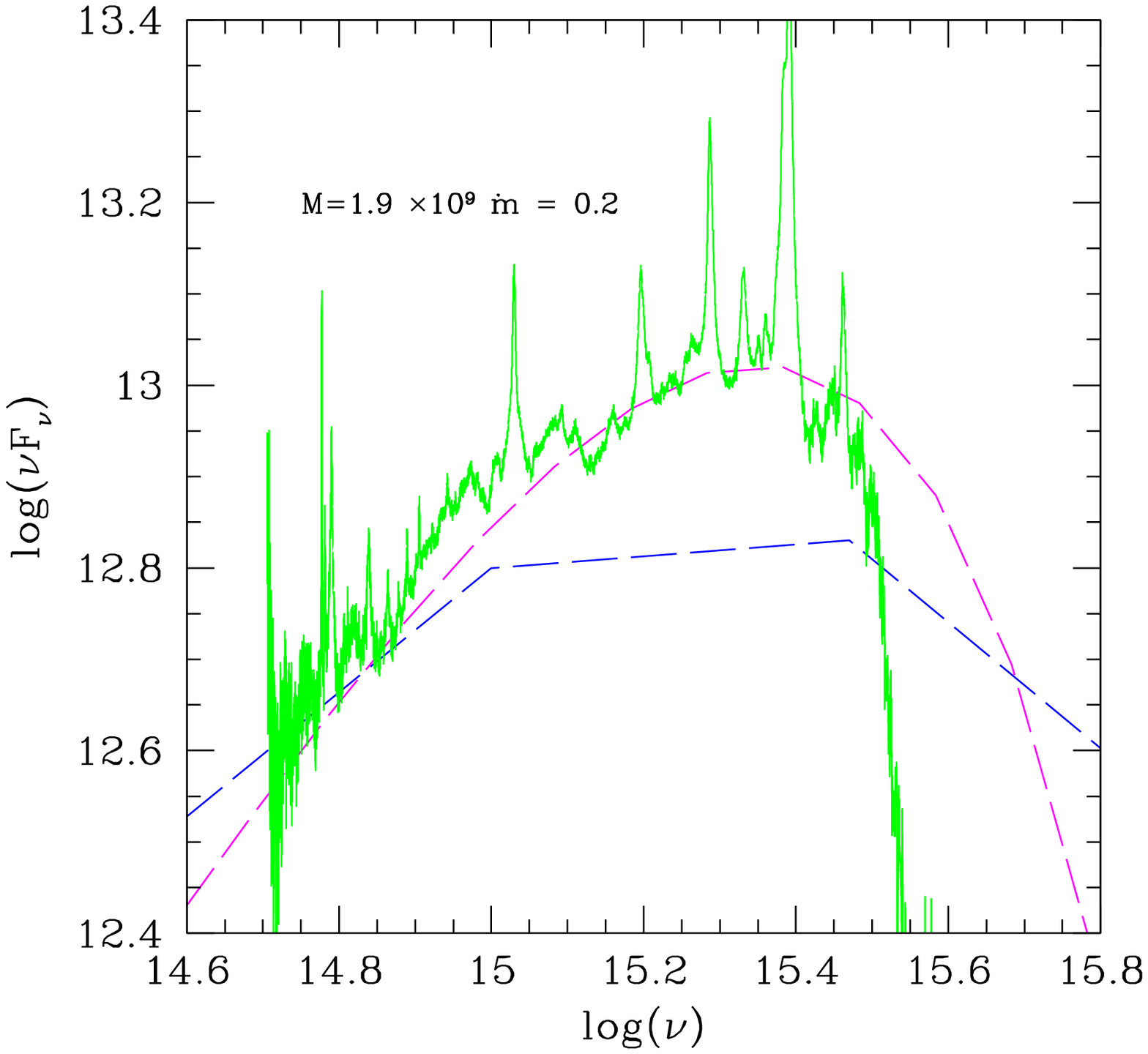}}
\hfil
\parbox{0.5\textwidth}{
\epsfxsize=0.45\textwidth
\vskip -1.7 true cm
\epsfbox[18 200 400 0]{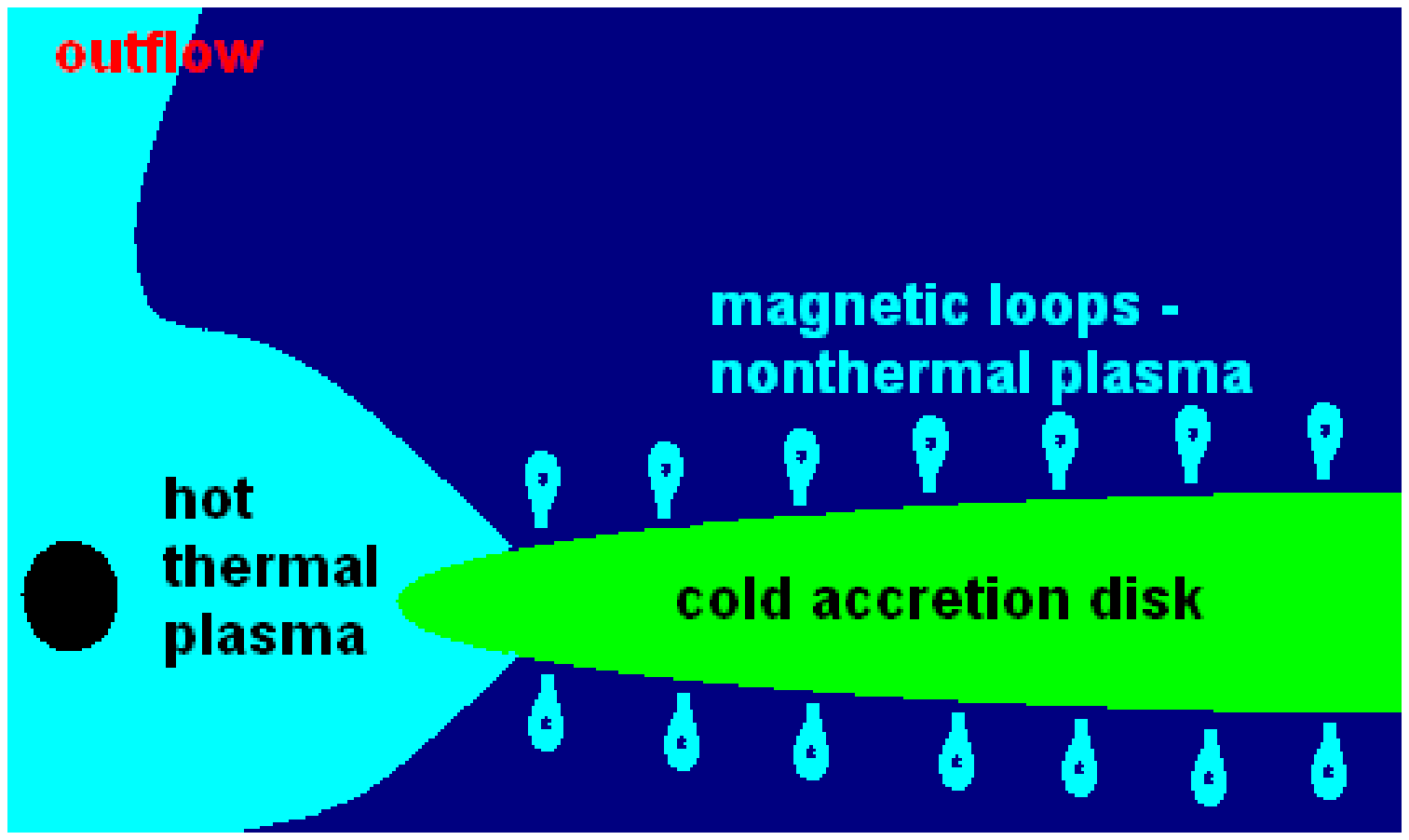}}
}
\caption{Left: Optical/UV composite spectrum of very bright quasars 
(Francis  et al. 1991, continuous line), of moderately bright quasars 
(Laor et al. 1997, long-dash line), together with an exemplary spectrum of
a simple Shakura-Sunyaev disk (short-dash line). Right: Schematic view 
of the accretion flow in a bright AGN.
\label{fig:geom}}
\end{figure}

The broad band spectrum clearly shows that the accretion flow is 
(at least) a two-phase medium. The profound optical/UV/soft X-ray bump (Big
Blue Bump) so characteristic for high $L/L_{Edd}$ sources is well
modeled as a thermal emission of a Keplerian, optically thick and geometrically
thin disk (see Fig.~\ref{fig:geom}). The hard X-ray emission must come from a
hot optically thin plasma in the vicinity of the disk. The IR emission is due
to reprocessing of a part of the radiation by circumnuclear dust, usually referred to
as a dusty-molecular torus. The presence of this torus is responsible for shielding
the central parts for highly inclined observers, as is the case for type 2 objects.
The torus is most probably clumpy, and quite possibly it is rather a kind of outflowing
dusty wind instead of a structure in the hydrostatic equilibrium.

The location and geometry of the hot plasma is uncertain. A plausible possibility is shown
in Fig.~\ref{fig:geom}. The disk is covered by magnetic loops emerging from its interior 
\cite{galeev}. The
mechanism behind this is the magneto-rotational instability (MRI) operating in the disk interior.
This instability is responsible for the disk viscosity, and the corresponding 
Shakura-Sunyaev parameter $\alpha$ is $\sim 0.01$, according to numerical simulations.
Occasionally, large loops emerge high above the disk surface \cite{MilStone00}, 
and the magnetic field 
reconnection results in formation of a flare (hard X-ray flash). Therefore, a corona 
similar to the solar corona forms above the disk. In the innermost part the
cold disk can be evaporated due to the interaction with such an active corona \cite{aga2000}. 
A flow towards the black hole proceeds through a
hot plasma phase \cite{SLE,ichimaru,narayan}, and a (perhaps significant) fraction of the plasma can be expelled out
in the process. In strongly radio-loud objects this outflow takes a form of a 
relativistic well-collimated jet but in radio-quiet objects the outflow is slower 
and uncollimated. The mechanism of the outflow is unknown.

Within this picture, we would expect the formation of non-thermal plasma in the coronal loops above the disk while in the hot phase the electrons would be predominantly thermal. The relative role of the
two media is determined by the disk trucation radius which in turn is determined by $L/L_{Edd}$. At
$L/L_{Edd} \sim 0.1 - 0.5 $ or more the disk is supposed to extend down to the marginally stable orbit. 

The geometry shown in Fig.~\ref{fig:geom} is not a unique possibility. Other accretion flow models include (i) the lamp-post model in which the disk always extends up to the marginally stable orbit and the high energy emission comes from a shock formed in the outflowing plasma and localized on the symmetry axis \cite{guy,miniutti} (ii) disk always extending down to the marginally stable orbit with magnetic flares above it \cite{galeev} (iii) accretion in the form of clumps of cold material embedded in a hot plasma \cite{collin96}. In those models the spectral differences between
high Eddington and low Eddington sources are less naturally explained.

\section{X-ray spectroscopy as a tool to study the plasma motion}

Given the lack of sufficient spatial resolution the key to the flow geometry 
lies in X-ray spectroscopy. First extremely useful results came from Ginga \cite{ginga12} and ASCA \cite{tanaka95} data, now Chandra and XMM-Newton offer still higher quality data. A number of blueshifted absorption lines was identified which allowed to measure the outflow velocity of the material, and measured a number of emission line profiles, including the famous iron K$\alpha$ line. 

\subsection{outflow}

Chandra and XMM-Newton observations confirmed the earlier findings that a partially ionized warm 
absorber exists in many Seyfert 1 and NLS1 galaxies \cite{Blustin}. Outflow velocities range from
hundreds km s$^{-1}$ in Seyfert 1 \cite{kaspi,kaastra} galaxies to a fraction of light speed in 
NLS1 \cite{chartas,pounds1211}. The distance of this outflowing material from the black hole is 
difficult to assign. Simple arguments based on the assumption of a roughly Keplerian velocity 
suggest that the slow outflow originates somewhere between the Broad Line Region and a Narrow 
Line Region, i.e. at a few parsecs from the nucleus, and the fast outflow originates closer in,
at distances of order of hundreds of Schwarzschild radii. The amount of outflowing material may be
quite large - fast outflows may actually carry out considerable fraction of the inflowing 
material \cite{kingpounds}. The estimates are difficult since a fraction of the material may be 
completely ionized and leave no direct signature in the soft X-ray spectrum. However, this 
material will scatter a fraction of the nuclear emission and may lead to modification of the 
optical spectrum of an AGN through the irradiation of the outer disk \cite{uni,loska}. 

\subsection{quasi-Keplerian motion}

\begin{figure}
\vskip -7.5 true cm
\parbox{\textwidth}{
\parbox{0.30\textwidth}{
\epsfxsize=0.25\textwidth
\vskip -1 true cm
\epsfbox[0 0 400 900]{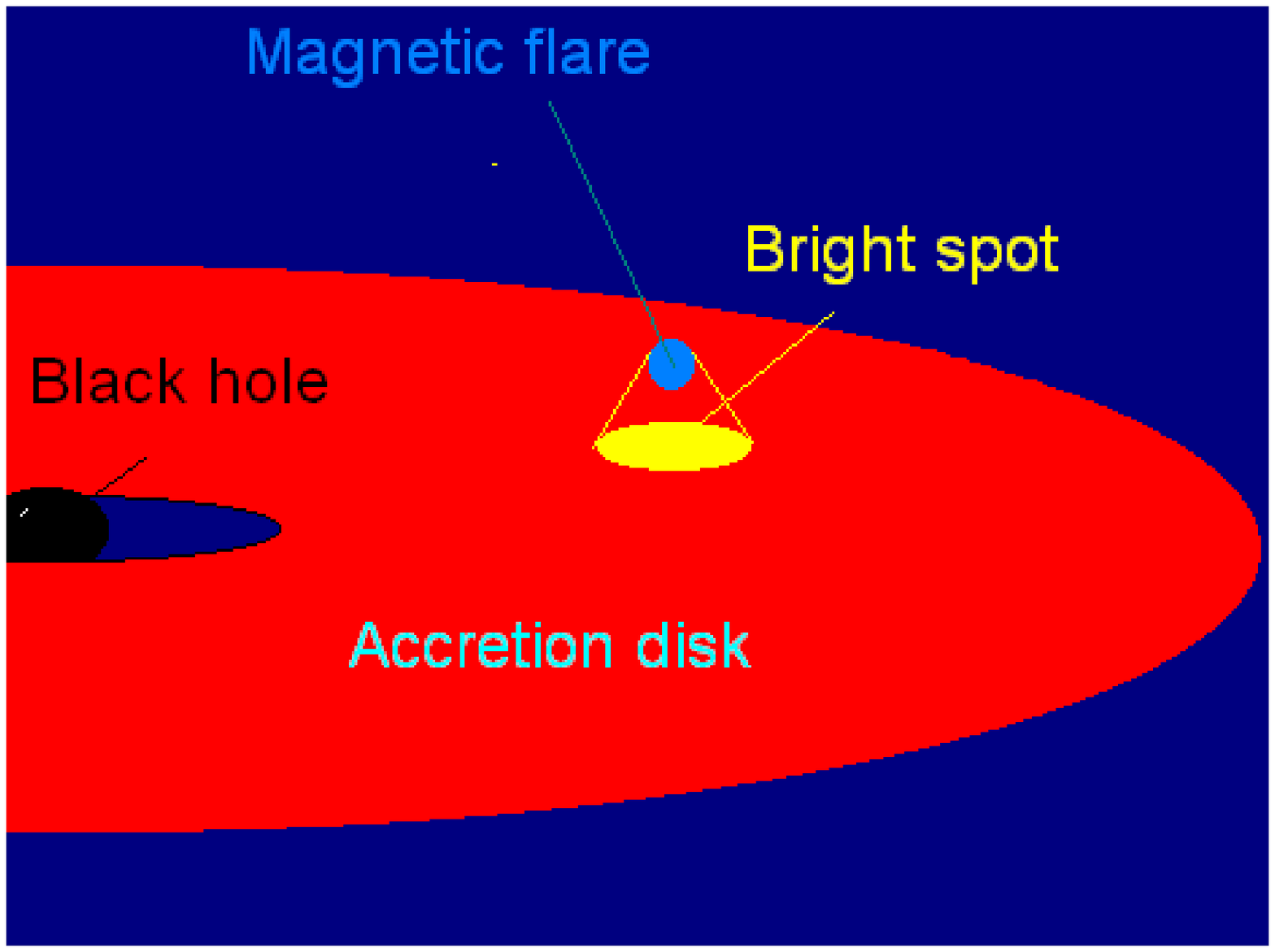}}
\hfil
\parbox{0.30\textwidth}{
\epsfxsize=0.25\textwidth
\vskip -3 true cm
\epsfbox[0 0 300 900]{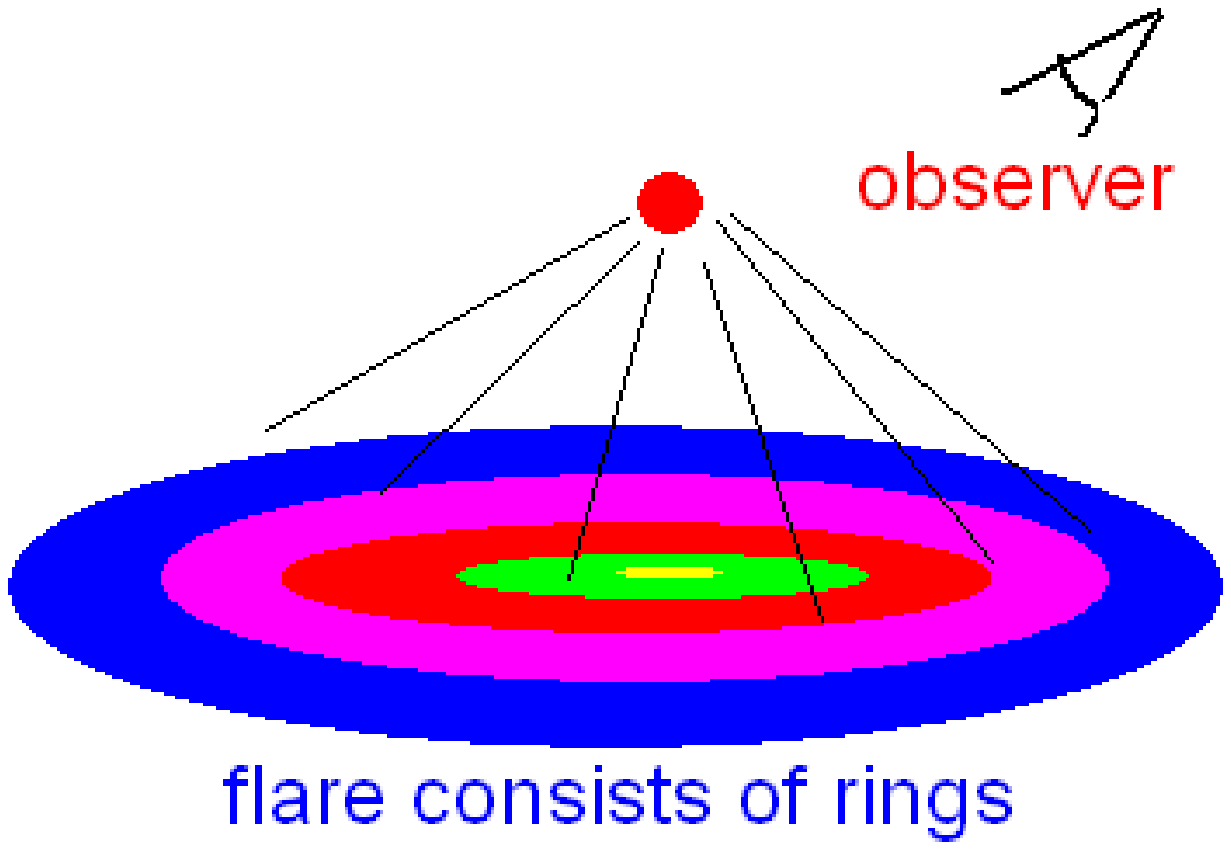}}
\hfil
\parbox{0.30\textwidth}{
\epsfxsize=0.25\textwidth
\vskip 2 true cm
\hskip -2 truecm
\epsfbox[0 0 300 900]{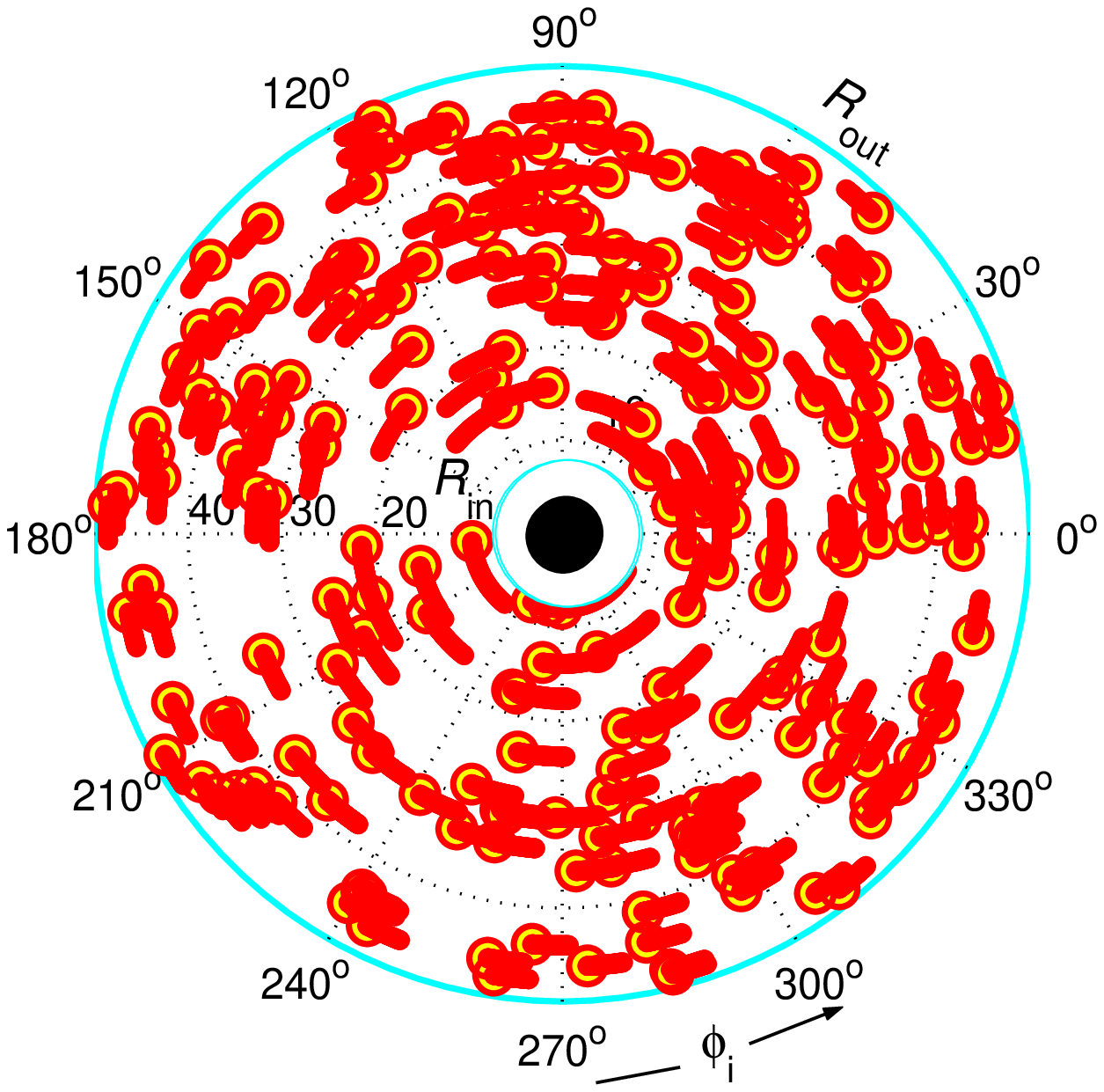}}
}
\parbox{\textwidth}{
\epsfxsize=0.35\textwidth
\center
\vskip -3 true cm
\epsfbox{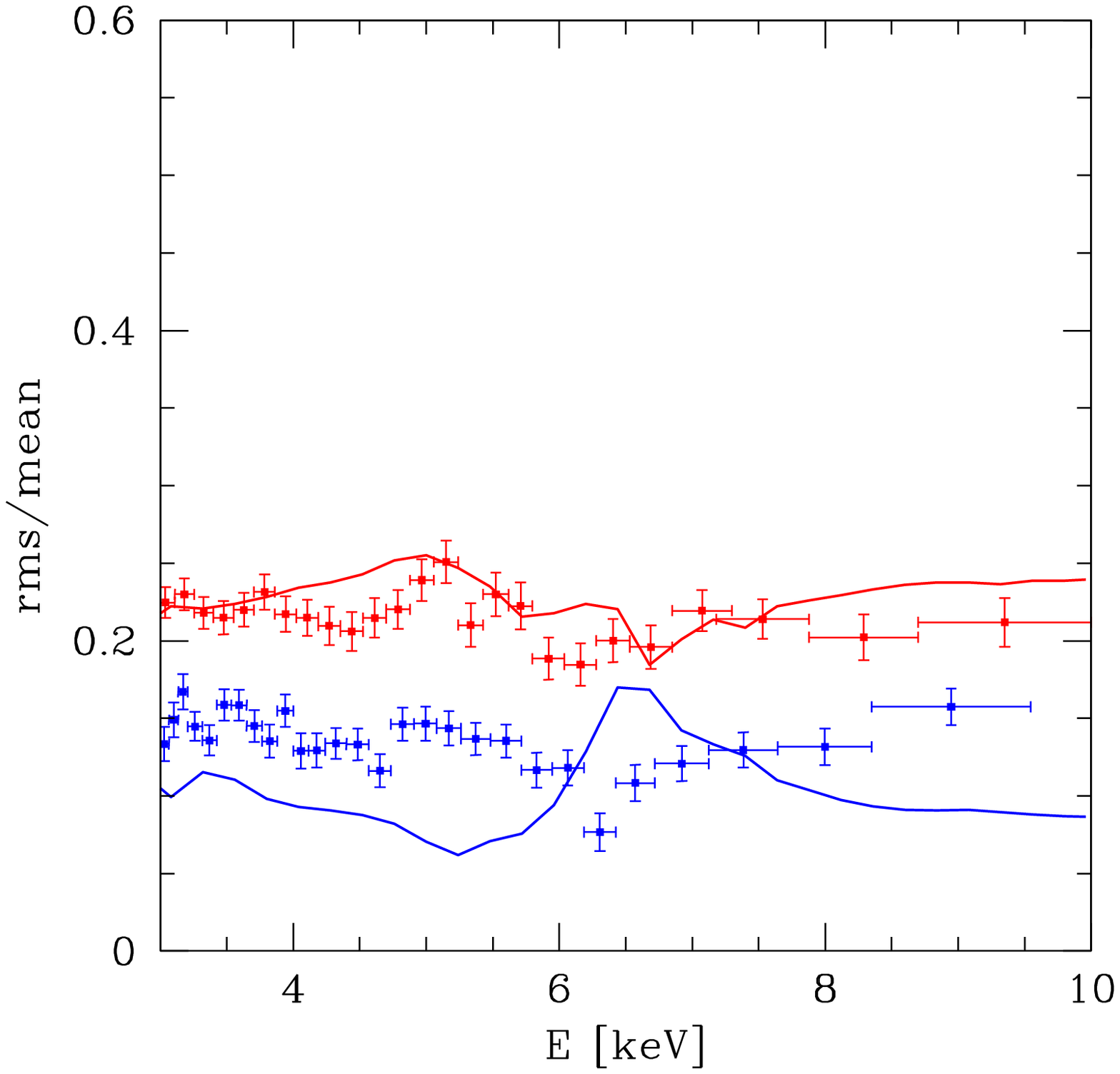}}
\caption{Model of the variability in MCG-6-15-30 in X-ray band. Each magnetic flare
creates a hot spot at the disk surface (left), the energy-dependent single spot emission 
is integrated assuming certain inclination angle of an observer, a time-dependent spot
distribution across the disk surface is generated (right), and the time-dependent
spectrum is calculated taking into account the relativistic effects. Finally, fractional
variability amplitude (upper curve) and point-to-point fractional variability amplitude
are obtained and compared to the observed one. 
\label{fig:flary}}
\end{figure}

Chandra and XMM-Newton confirmed that in some objects the very broad emission 
lines (iron line and soft X-ray lines) are seen, with the shape well 
represented as the effect of the relativistic smearing due to the Keplerian 
motion of the emitting material located very close to the black hole \cite{reviron}.
Some observations indicate that the inner radius of the disk increases when the source
becomes fainter \cite{matt}.

However, the observed variability of the iron line is not well understood and remains a
major puzzle \cite{longinotti,bianchi}. Models require some fine tuning in order to reproduce the
observed trends.

Models which explore full observational information have even more difficulties. We attempted
to reproduce both the fractional variability amplitude and point-to-point fractional variability
amplitude for MCG-6-15-30. The model assumed that magnetic flares are randomly distributed 
above the disk surface. The flare flux, flare duration and the probability of a flare at a given 
radius were assumed to have a power law dependence on the radius. Black hole was assumed to
rotate, disk was assumed to extend down to the marginally stable orbit appropriate for an adopted
Kerr parameter. Each flare created a hot spot on the disk surface by irradiation and the 
reflection (including iron line and other soft lines) was calculated using the code 
{\sc titan/noar} \cite{titan} and taking into account that the incident flux depends on the 
distance from the flare center. Flare was assumed to be short-lasting so the vertical structure 
of the disk was taken from an unilluminated model and the disk expansion was neglected. Predicted 
spectrum was integrated in specific time bins, as in the
data, all relativistic effects were included using the code {\sc ky} \cite{ky}.

Within this scenario, we could not find a parameter range which would reproduce the shape of both
fractional variability functions. Either still the model is too simple (we did not consider
the dependence of the shape of the reprocessed radiation on the radius) or an important element
is still missing. There is some evidence that variations in the outflowing warm absorber can
contribute significantly. It was even suggested that warm absorber is mostly responsible for the
observed shape of the spectrum in soft X-ray band \cite{gierdone}, and the broad iron line is
partially an artefact \cite{pounds05}.

\section{Shocks and magnetic field reconnections as a heating mechanism of the hot plasma}

Although the geometry of the hot material is still under discussion, some mechanisms of 
plasma heating are needed to explain the observed X-ray emission. 

Models that can apply to magnetic flares above the disk surface were discussed in numerous 
papers \cite{roma,liu}.

In ADAF type solutions most of the gravitational energy is used to heat ions, and subsequently
directed towards electrons through Coulomb interaction \cite{SLE,macha}. However, a fraction
of energy will be inevitably used to heat electrons directly \cite{gena1} thus limiting the low
radiative efficiency of the flow. 

Several other aspects were also discussed, like disk-corona coupling \cite{kun}, disk evaporation
and the ion irradiation of the disk \cite{spruit} but the picture
is far from being complete. 

Observationally, very interesting results were obtained in the
context of a galactic source (microquasar) GRS~1915+105 \cite{febel} and they may apply to
AGN when timescales are corrected by the black hole mass ratio. Methods, like Fourier-resolved
spectroscopy, successfully applied to galactic sources \cite{zycki2003}, may be also helpful although the
AGN data at present are hardly of the appropriate quality. 

Therefore, observational determination of the gamma-ray emission from radio-quiet objects

\section*{Acknowledgments}
We thank Aneta Siemiginowska and Piotr \. Zycki for very helpful discussions.
Part of this work was supported by the grant PBZ-KBN-054/P03/2001 
of the Polish
State Committee for Scientific Research, and by the Laboratoire Europe\' en Associ\' e Astrophysique 
Pologne-France.


\section*{References}
\begin{thebibliography}{99}

\bibitem{woo} J.-H. Woo and C.M. Urry, \Journal{\apj}{579}{530}{2002}
\bibitem{highz} X. Fan et al., \Journal{\aj}{128}{515}{2004}
\bibitem{sdss} http://www.sdss.org/
\bibitem{wandel}A. Wandel, Proc. of IAU Symp. 222, Black Holes, Stars and ISM in Galactic 
   Nuclei, astro-ph/0407399 (2004)
\bibitem{tran} H.D. Tran, \Journal{\apj}{583}{632}{20003}
\bibitem{bachev} R. Bachev et al., \Journal{\apj}{617}{171}{2004}
\bibitem{BLR} B. Czerny, A. R\' o\. za\' nska and J. Kuraszkiewicz, \Journal{\aa}{428}{39}{2004}
\bibitem{elvis94} M.Elvis et al., \Journal{\apjs}{95}{413}{1994}
\bibitem{laor} A. Laor et al., \Journal{\apj}{477}{93}{1997}
\bibitem{markowitz} A. Markowitz et al, \Journal{\apj}{593}{96}{2003}
\bibitem{optical} T. Totani et al., \Journal{\apj}{621}{L9}{2005}
\bibitem{giveon} U. Giveon et al., \Journal{\mn}{306}{637}{1999}
\bibitem{Zdzia-OSSE} A.A. Zdziarski, J. Poutanen and W.N. Johnson, \Journal{\apj}{542}{703}{2000}
\bibitem{deluit}S. Deluit and T.J.-L. Courvoisier, \Journal{\aa}{399}{77}{2003}
\bibitem{background}N. Menci, F. Fiore , G.C. Perola, and A. Cavaliere, \Journal{\apj}{606}{58}{2004}
\bibitem{sobo1} M. Sobolewska, A. Siemiginowska and P.T. \. Zycki, \Journal{\apj}{608}{80}{2004}
\bibitem{sobo2} M. Sobolewska, A. Siemiginowska and P.T. \. Zycki, \Journal{\apj}{617}{102}{2004}
\bibitem{ghisellini} G. Ghisellini, R. Svensson, Proceedings of the NATO Advanced Research Workshop Physical Processes in Hot Cosmic Plasmas, held in Vulcano, Sicily, Italy, May 29-June 2, 1989. Eds, Wolfgang Brinkmann, Andrew C. Fabian, Franco Giovannelli; Publisher, Kluwer Academic Publishers, Dordrecht, The Netherlands, Boston, MA, (1990)
\bibitem{galeev} A.A. Galeev, R. Rosner and G.S. Vayana, \Journal{\apj}{229}{318}{1979}
\bibitem{MilStone00} K.A. Miller and J.M. Stone, \Journal{\apj}{534}{398}{2000}
\bibitem{aga2000} A. R\' o\. za\' nska and B. Czerny, \Journal{\aa}{360}{1170}{2000} 
\bibitem{SLE} S.L. Shapiro, A.P. Lightman and D.M. Eardley, \Journal{\apj}{204}{187}{1976}
\bibitem{ichimaru} S. Ichimaru, \Journal{\apj}{214}{840}{1977}
\bibitem{narayan} R. Narayan and I. Yi, \Journal{\apj}{428}{L13}{1994}
\bibitem{guy} G. Henri and G. Pelletier, \Journal{\apj}{383}{L7}{1991} 
\bibitem{miniutti} G. Miniutti and A.C. Fabian, \Journal{\mn}{349}{1435}{2004} 
\bibitem{collin96} S. Collin-Souffrin, B. Czerny, A.-M. Dumont, P.T. \. Zycki, \Journal{\aa}{314}{393}{1996}
\bibitem{ginga12} K.A. Pounds, K. Nandra, G.C. Stewart, I.M. George, A.C. Fabian, \Journal{\nat}{344}{132}{1990}
\bibitem{tanaka95} Y. Tanaka et al., \Journal{\nat}{375}{659}{1995}
\bibitem{Blustin}A.J. Blustin, M.J. Page, S.V. Fuerst, G. Branduardi-Raymont and C.E. Ashton, \Journal{\aa}{431}{111}{2005}
\bibitem{kaspi} S. Kaspi et al., \Journal{\apj}{554}{216}{2001}
\bibitem{kaastra} J.S. Kastra et al., \Journal{\aa}{386}{427}{2002}
\bibitem{chartas} G. Chartas, W.N. Brandt, S.C. Gallagher, and G.P. Garmire, \Journal{\apj}{279}{169}{2002}
\bibitem{pounds1211} K.A. Pounds, A.R. King, K.L. Page and P.T. O'Brien, \Journal{\mn}{346}{102}{2003}
\bibitem{kingpounds} A.R. King and K.A. Pounds, \Journal{\mn}{345}{657}{2003}
\bibitem{uni}B. Czerny et al., \Journal{\aa}{412}{317}{2003}
\bibitem{loska} Z. Loska, B. Czerny and R. Szczerba, \Journal{\mn}{355}{1080}{2004}
\bibitem{reviron} C.S. Reynolds and M.A. Nowak, \Journal{\phr}{377}{389}{2003}
\bibitem{matt} G. Matt et al., astro-ph/0502323 (2005)
\bibitem{longinotti} A.L. Longinotti, K. Nandra, P.O. Petrucci, P.M. O'Neil, \Journal{\mn}{355}{929}{2004}
\bibitem{bianchi} S. Bianchi et al., \Journal{\aa}{422}{65}{2004}
\bibitem{titan} A.-M. Dumont, A. Abrassart and S. Collin, \Journal{\aa}{357}{823}{2000}
\bibitem{ky} M. Dovciak, V. Karas and T. Yaqoob, \Journal{\apjs}{153}{205}{2004}
\bibitem{gierdone} M. Gierli\' nski and C. Done, \Journal{\mn}{349}{L7}{2004}
\bibitem{pounds05} K.A. Pounds, astro-ph/0505447 (2005)
\bibitem{roma} D.A. Larrabee, R.V.E. Lovelace  and M.M. Romanova, \Journal{\apj}{586}{72}{2003}
\bibitem{liu} B.F. Liu , S. Mineshige , and K. Ohsuga, \Journal{\apj}{587}{571}{2003}
\bibitem{macha} R. Narayan, R. Mahadevan and E. Quataert, E. 1999, in The Theory of Black Hole Accretion Discs, 
ed. M. A. Abramowicz, G. Bjornsson, and J. E. Pringle (Cambridge: Cambridge Univ. Press)
\bibitem{gena1} G.S. Bisnovatyi-Kogan and R.V.E. Lovelace, \Journal{\apj}{529}{978}{2000}
\bibitem{kun} Z. Kuncic and G.V. Bicknell, \Journal{\apj}{616}{669}{2004}
\bibitem{spruit} C. P. Dullemond and H.C. Spruit, \Journal{\aa}{434}{415}{2005}
\bibitem{febel} R. Fender and T. Belloni, \Journal{\araa}{42}{317}{2004}
\bibitem{zycki2003} P.T. \. Zycki, \Journal{\mn}{340}{639}{2003}


\end{thebibliography}
\end{document}